\documentclass[longauth]{aa} 
\usepackage{graphicx}
\usepackage{natbib}
\usepackage{txfonts}
\newcommand{\xirppi}{$\xi(r_p,\pi)$}

\newcommand{\wprp}{$w_p(r_p)$}
\newcommand{\ro}{$r_0$}
\newcommand{\gam}{$\gamma$}

\def\omnow{\Omega_{\rm m,0}}

\def\ovnow{\Omega_{\rm \Lambda,0}}

\def\simge{\mathrel{%
   \rlap{\raise 0.511ex \hbox{$>$}}{\lower 0.511ex \hbox{$\sim$}}}}
\def\simle{\mathrel{
   \rlap{\raise 0.511ex \hbox{$<$}}{\lower 0.511ex \hbox{$\sim$}}}}

\DeclareTextSymbol{\degre}{T1}{6}
\DeclareTextSymbol{\degre}{OT1}{23}

\begin{document}
   \title{The VIMOS-VLT Deep Survey (VVDS)}

   \subtitle{The dependence of clustering on galaxy stellar mass at $z\sim 1$
     \thanks{based on data
       obtained with the European Southern Observatory Very Large
       Telescope, Paranal, Chile, program 070.A-9007(A), and on data
       obtained at the Canada-France-Hawaii Telescope, operated by
       the CNRS of France, CNRC in Canada and the University of Hawaii.
       Based on observations obtained with MegaPrime/MegaCam, a joint  
       project of CFHT and CEA/DAPNIA, at the Canada-France-Hawaii Telescope  
       (CFHT) which is operated by the National Research Council (NRC) of  
       Canada, the Institut National des Science de l'Univers of the Centre  
       National de la Recherche Scientifique (CNRS) of France, and the  
       University of Hawaii. This work is based in part on data products  
       produced at TERAPIX and the Canadian Astronomy Data Centre as part of  
       the Canada-France-Hawaii Telescope Legacy Survey, a collaborative  
       project of NRC and CNRS.
     }
   }

   \author{
          B. Meneux \inst{1,2}
     \and L. Guzzo \inst{3,2,4,5}
     \and B. Garilli \inst{1}
     \and O. Le F\`evre \inst{6}
     \and A. Pollo \inst{6,7}
     \and J. Blaizot \inst{4}
     \and G. De Lucia \inst{4}
     \and M. Bolzonella  \inst{8} 
     \and F. Lamareille \inst{8}
     \and L. Pozzetti    \inst{8}
     \and A. Cappi    \inst{8}
     \and A. Iovino \inst{9}
     \and C. Marinoni \inst{10}
     \and H.J. McCracken \inst{11,12}
     \and S. de la Torre \inst{6}
     \and D. Bottini \inst{1}
     \and V. Le Brun \inst{6}
     \and D. Maccagni \inst{1}
     \and J.P. Picat \inst{13}
     \and R. Scaramella \inst{14,15}
     \and M. Scodeggio \inst{1}
     \and L. Tresse \inst{6}
     \and G. Vettolani \inst{14}
     \and A. Zanichelli \inst{14}
     \and U. Abbas \inst{6}
     \and C. Adami \inst{6}
     \and S. Arnouts \inst{6}
     \and S. Bardelli  \inst{8}
     \and A. Bongiorno \inst{16}
     \and S. Charlot \inst{4,11}
     \and P. Ciliegi    \inst{8}  
     \and T. Contini \inst{13}
     \and O. Cucciati \inst{9,17}
     \and S. Foucaud \inst{18}
     \and P. Franzetti \inst{1}
     \and I. Gavignaud \inst{19}
     \and O. Ilbert \inst{20}
     \and B. Marano     \inst{16}  
     \and A. Mazure \inst{6}
     \and R. Merighi   \inst{8} 
     \and S. Paltani \inst{21,22}
     \and R. Pell\`o \inst{13}
     \and M. Radovich \inst{23}
     \and D. Vergani \inst{1}
     \and G. Zamorani \inst{8} 
     \and E. Zucca    \inst{8}
   }

   \offprints{B. Meneux bmeneux@mpe.mpg.de}

   \institute{
INAF-IASF - via Bassini 15, I-20133, Milano, Italy
\and
Max Planck Institut fuer extraterrestrische Physik, D-85741, Garching, Germany
\and
INAF-Osservatorio Astronomico di Brera - Via Bianchi 46, I-23807 Merate (LC), Italy 
\and
Max Planck Institut fuer Astrophysik, D-85741, Garching, Germany
\and
European Southern Observatory, D-85741, Garching, Germany
\and
Laboratoire d'Astrophysique de Marseille, UMR 6110 CNRS-Universit\'e de Provence,  BP8, 13376 Marseille Cedex 12, France
\and
Astronomical Observatory of the Jagiellonian University, ul Orla 171, 30-244 Krak{\'o}w, Poland
\and
INAF-Osservatorio Astronomico di Bologna - Via Ranzani,1, I-40127, Bologna, Italy 
\and
INAF-Osservatorio Astronomico di Brera - Via Brera 28, Milan, Italy
\and
Centre de Physique Th\'eorique, UMR 6207 CNRS-Universit\'e de Provence, F-13288 Marseille France
\and
Institut d'Astrophysique de Paris, UMR 7095, 98 bis Bvd Arago, 75014 Paris, France
\and
Observatoire de Paris, LERMA, 61 Avenue de l'Observatoire, 75014 Paris, France
\and
Laboratoire d'Astrophysique de Toulouse/Tabres (UMR5572), CNRS, 
Universit\'e Paul Sabatier - Toulouse III, Observatoire Midi-Pyr\'en\'ees, 14 
av. E. Belin, F-31400 Toulouse (France)
\and
IRA-INAF - Via Gobetti,101, I-40129, Bologna, Italy
\and
INAF-Osservatorio Astronomico di Roma - Via di Frascati 33, I-00040, Monte
Porzio Catone, Italy
\and
Universit\`a di Bologna, Dipartimento di Astronomia - Via Ranzani,1,
I-40127, Bologna, Italy
\and
Universit\'a di Milano-Bicocca, Dipartimento di Fisica - Piazza delle Scienze, 3, I-20126 Milano, Italy
\and	
School of Physics \& Astronomy, University of Nottingham, University Park, Nottingham, NG72RD, UK
\and
Astrophysical Institute Potsdam, An der Sternwarte 16, D-14482
Potsdam, Germany
\and
Institute for Astronomy, 2680 Woodlawn Dr., University of Hawaii,
Honolulu, Hawaii, 96822
\and
Integral Science Data Centre, ch. d'\'Ecogia 16, CH-1290 Versoix
\and
Geneva Observatory, ch. des Maillettes 51, CH-1290 Sauverny, Switzerland
\and
INAF-Osservatorio Astronomico di Capodimonte - Via Moiariello 16, I-80131,
Napoli, Italy
   }

   \date{Received --; accepted --}

   \abstract
   {}
   {We present a measurement of the dependence of galaxy
     clustering on galaxy stellar mass at redshift $z\sim0.9$,
     based on the first-epoch data from the VVDS-Deep survey.
   }
   {Concentrating on the redshift interval $0.5<z<1.2$, we measure
     the projected correlation function, \wprp, within
     mass-selected sub-samples covering the range $\sim10^{9}$
     and $\sim10^{11}~M_\odot$. We explore and quantify in detail
     the observational selection biases due to the flux-limited
     nature of the survey, both from the data themselves and using
     a suite of realistic mock samples constructed by coupling the
     Millennium Simulation to semi-analytic models. We
     identify the range of masses within which our main
     conclusions are robust against these effects.  
     Serious incompleteness in mass is present below
     $\log(M/M_\odot)=9.5$, with about two thirds of the galaxies
     in the range $9<\log(M/M_\odot)<9.5$ that are lost due to
     their low luminosity and high mass-to-light ratio.  However,
     the sample is expected to be 100\% complete in
     mass above $\log(M/M_\odot)=10$.  
   }
   {We present the first direct evidence for a dependence of
     clustering on the galaxy stellar mass at a redshift as high
     as $z\sim0.85$.  We quantify this by fitting the projected
     function \wprp\ with a power-law model.
     The clustering length increases from
     $r_0=2.76_{-0.15}^{+0.17}~h^{-1}~Mpc$ for galaxies with mass
     $M>10^{9}~M_\odot$ to 
     $r_0=4.28_{-0.45}^{+0.43}~h^{-1}~Mpc$ when only the most massive
     ($M>10^{10.5}~M_\odot$) are considered. 
     At the same time, we observe a significant increase in the slope, which
     over the same range of masses, changes from 
     $\gamma=1.67_{-0.07}^{+0.08}$ to
     $\gamma=2.28_{-0.27}^{+0.28}$. 
     
     Comparison to the SDSS measurements at $z\sim0.15$ shows that the 
     evolution of \wprp\ is significant for samples of
     galaxies with $M<10^{10.5}~M_\odot$, while it is negligible
     for more massive objects. Considering the growth of
     structure, this implies that the linear bias $b_L$ of the
     most massive galaxies evolves more rapidly between these two
     cosmic epochs.
     We quantify this effect by computing
     the value of $b_L$ from the SDSS and VVDS clustering amplitudes
     and find that $b_L$ decreases from $1.5\pm0.2$ at $z\sim0.85$ to
     $1.33\pm0.03$ at $z\sim0.15$, for the most massive galaxies,
     while it remains virtually constant ($b_L\sim1.3$) for the
     remaining population.
     Qualitatively, this is the kind of scenario expected for the clustering
     of dark-matter halos as a function of their total mass and redshift.
     Our result seems therefore to indicate that galaxies with the largest
     stellar mass today were originally central objects of the
     most massive dark-matter halos at earlier times, whose
     distribution was strongly biased with respect to the overall
     mass density field.
   }
   {}
   
       
   \authorrunning{Meneux, B., et al.}
   \titlerunning{Galaxy clustering as a function of stellar mass}

   \maketitle
%

\section{Introduction}

In the currently accepted scenario, galaxies are thought to form
within extended dark matter halos \citep{whiterees1978}, which grow
through subsequent mergers in a hierarchical fashion.   A major
challenge in testing this general picture is to connect the observable 
properties of galaxies to those of the 
dark-matter halos in which they are embedded, as
predicted, e.g., by large n-body simulations
\citep[e.g.][]{springel2006}. 

At the current epoch, the observed clustering of galaxies is found to
depend significantly on their specific properties, such as
luminosity
\citep{hamilton1988, iovino1993, maurogordato1991, benoist1996, guzzo2000,
  norberg2001, norberg2002, zehavi2005}, 
color or spectral type \citep{willmer1998,norberg2002,zehavi2002},
morphology \citep{davisgeller1976,giovanelli1986,guzzo1997}
and stellar mass \citep{li2006}.
Similar conclusions are drawn at high redshift from deep galaxy surveys
\citep{coil2006, pollo2006, phleps2000, phleps2006, meneux2006, daddi2003}.

It is somewhat natural to expect that quantities such as luminosity and, in
particular, stellar mass of galaxies should in some way be related to the mass
of the dark-matter halo. At the time of observation, this is strictly true
only for isolated galaxies \citep{skibba2007}, but moving back along the
evolutionary path of the galaxy one can always find a time when this was true,
before the galaxy and its halo were accreted by a larger dark-matter structure.

Recent theoretical works seem to indicate that, indeed, there exists a fairly
direct relationship between global galaxy properties (e.g.~their stellar or
total baryonic mass, or their luminosity) and the halo mass, {\it before} it
is accreted by a larger dark-matter halo
\citep[$M_{\rm  infall}$,][]{conroy2006, wang2006, wang2007}. In practice,
this represents the mass of the dark matter halo of the galaxy, at a time when
this was still the dominant central object of its own halo.
Sensibly enough, it is $M_{\rm infall}$ which is defining the global
properties of the baryonic component. 
This complex relation between galaxies and the (sub)halos in which they
reside  \citep{springel2001,gao2004} offers a novel
way to interpret the observed clustering properties of galaxies.
Theoretically, it provides also more direct physical insight into the
origin of the observed relationships, with respect to standard Halo
Occupation Distribution (HOD) models \citep[see][]{cooraysheth2002}.

Observationally, stellar mass has become a quantity measured with
increasing accuracy, thanks to surveys with multi-wavelength
photometry, extending to the near-infrared
\citep[e.g.][]{rettura2006}, although some uncertainties related to
the detailed modelling of stellar evolution remain
\citep{pozzetti2007}. This makes studies of clustering as a
function of stellar mass possible for large statistical
samples. \citet{li2006} have measured the dependence of clustering on
stellar mass (and other properties) in the local Universe, using a
sample of $\sim200,000$ galaxies drawn from the Sloan Digital Sky
Survey (SDSS). They find, not surprisingly, that galaxies of larger
mass are more clustered than low-mass ones, with the effect increasing
above the characteristic knee value $M^*$ of the Schechter mass
function.  These measurements represent a high-quality reference to
which high-redshift measurements can be compared to test evolution.

In studying the clustering of galaxies at different redshifts, a
selection based on mass should also reduce the typical ``progenitor''
biases intrinsic when using flux-limited samples to study evolution of
large-scale structure.  Given the strong luminosity evolution between
redshift zero and one, and its dependence on galaxy spectral type \citep{zucca2006},
stellar mass should guarantee a more stable parameter on the basis of
which comparing galaxies of possibly similar sort.  If stellar mass
does not evolve appreciably below $z\sim1$ and the merger rate is
also negligible in the same range (which might not be true, see \citet{bell2006}),
the growth of clustering observed for objects of fixed
mass should reflect directly the evolution of the clustering of parent
dark-matter halos.  In fact, the stellar content of galaxies is also
thought to evolve between $z=1$ and now, but this should correspond to
a maximum increase in mass of a factor of 2 \citep{arnouts2007}. A further
complication to this idealized scenario, comes from the fact that galaxy
surveys are inevitably flux-limited. Therefore, when extracting mass-selected
samples, especially at high redshift, particular care has to be taken in
considering selection effects that might bias the final results,
essentially loosing high mass-to-light ratio galaxies at low mass
(see \citet{li2006} for discussion and correction of these effects at
$z\sim0$).

Using the VIMOS-VLT Deep Survey (VVDS), we have the possibility to
investigate for the first time the evolution with redshift of the
clustering in a mass-selected galaxy sample. On one side, the VVDS-Deep
(see \ref{sec:vvdsdata} below), gives us the 
opportunity to compute accurate stellar masses, by virtue of its
multi-band coverage that extends to the near-infrared (see
\citet{pozzetti2007} for detailed discussion on the accuracy and
intrinsic limitations of mass estimates). On the other hand, it
provides us with an unprecedented area coverage for its depth
($I_{AB}\le 24$), thus allowing the measurement of spatial statistics
like the first (mean density) and second moments (two-point
correlation function) of the galaxy distribution.

Closely related to this study is the work of \citet{pollo2006}, where the
dependence of clustering on galaxy $B-band$ luminosity at $<z>\simeq 1$ has
been measured using the VVDS.  
They show that at this epoch, luminous galaxies show not only a higher clustering than
faint objects (similar to what is observed in the local Universe), but that
their correlation function is much steeper than local counterparts. Since the
$B$-band is sensitive to emission from young stars, comparison to the results
presented in this paper is of interest to understand how star formation and
stellar mass assembly relate to the evolution of the local environment (and
vice-versa).

In the present paper, we show the first measurement of the two point
correlation function as a function of stellar mass at redshift $z\sim0.85$, for
galaxies more massive than $10^9~M_\odot$.
The paper is organized as follows.
We introduce the data, the way stellar masses are derived and our galaxy
samples in Sect.~\ref{sec:data}.
We present the mock catalogues used throughout this paper in
Sect.~\ref{sec:mocks}.
The projected correlation function and its computation is introduced in
Sect.~\ref{sec:xi_technic}.
We discuss the stellar mass completeness of our sample and its effect on
the measurement of clustering properties in Sect.~\ref{sec:mass_comp}.
Results are presented in Sect.~\ref{sec:results}.
Summary and conclusions are in Sect.~\ref{sec:conclusion}.

Throughout this paper we use a Concordance Cosmology with $\Omega_m =
0.3$ and $\Omega_{\Lambda} = 0.7$. The Hubble constant is normally
parametrised via $h=H_0/100$ to ease comparison with previous works.
Stellar masses are quoted in unit of $h=1$.
All correlation length values are quoted in comoving coordinates. 


\section{Data}
\label{sec:data}

\subsection{The VVDS survey}
\label{sec:vvdsdata}

The VIMOS-VLT Deep Survey (VVDS) is performed with the VIMOS
multi-object spectrograph at the ESO Very Large Telescope
\citep{lefevre2003_spie} and complemented with multi-color BVRIJK imaging data
obtained at the CFHT and NTT telescopes \citep{hjmcc2003, lefevre2004_im,
iovino2005, radovich2003}.
For this work, we use the so-called ``first epoch''
data, collected in the F02 ``VVDS-Deep'' field.  This is a purely
magnitude limited sample to $I_{AB}=24$, covering an area of $0.49$
square degrees with a mean sampling of $\sim23\%$.  Considering only
galaxies with secure ($>80\%$ confidence) measurements, this sample
includes $6530$ redshifts, with an {\it rms} accuracy (estimated from
repeated observations) of
275 km s$^{-1}$.  This is the same data set used in all 
previous clustering analyses of the VVDS. Details about observations,
data reduction, redshift measurement and quality assessment can be
found in \citet{lefevre2005_vvds}.

\subsection{Mass-limited sub-samples}

Stellar masses for all galaxies in the VVDS catalogue were estimated
by fitting their Spectral Energy Distribution (SED), as sampled by the
VVDS multi-band photometry, with a library of stellar population
models based on \citet{bc2003}. Two different histories of star
formation were used, a smooth one and a complex one with secondary
bursts to derive two set of galaxy stellar mass.
The initial mass function of \citet{chabrier2003} was adopted
in each case. The resulting typical error on stellar mass
determination is $\sim$0.1 dex (depending on redshift and stellar
mass).  The reader is referred to \citet{pozzetti2007}
for a full description of the methodology used to derive the stellar
masses and for a discussion of their robustness and intrinsic errors.
In the next sections, we shall use stellar masses computed using a
complex star formation history only. In practice, we found that the
derived clustering properties are not sensitive to the details of the
method used.

To measure the dependence of clustering on stellar mass at  $z\sim1$, 
similarly to \citet{pollo2006}, we have restricted the VVDS-Deep data
to the redshift interval $[0.5-1.2]$.  In this range, containing 4285
redshifts, we have constructed a set of mass-limited sub-samples, as
defined in Tab.~\ref{tab:prop}.  These include four {\it differentially} binned
samples, D1, D2, D3 and D4, with narrow mass ranges corresponding
respectively to $\log(M/M_\odot)$=[9.0-9.5], [9.5-10.0], [10.0-10.5],
and [10.5-11.0].  We also consider four {\it integrated} samples, I1,
I2, I3 and I4, including galaxies with mass larger than a given
limit.
While the former are useful for direct comparison to local SDSS
measurements \citep{li2006}, results from the integral samples are
more robust, given their larger size.
As we shall discuss in detail in Sect.~\ref{sec:mass_comp}, the
lower-mass samples are affected, to different degrees, by incompleteness due
to the primary selection in observed flux, and the scatter in the
mass-luminosity relation. A large part of the following analysis will be
dedicated to understanding this incompleteness and its effect on the
measured clustering. These tests make extensive use of numerical
simulations, described in the following section.

\begin{table}
  \caption{Description of the 8 VVDS samples defined in the redshift
    range z=[0.5-1.2]
  } 
  \label{tab:prop}
  \centering
  \begin{tabular}{c c c c c}
    \hline\hline
Sample &\multicolumn{2}{c}{Stellar mass  ($\log(M/M_\odot)$)} & Mean     & Number of  \\
       &            range       &          median           & redshift  & galaxies   \\
    \hline
D1     &    $\ \ 9.0 - \ \ 9.5$ &        \ \ 9.25           &  0.83     &    1221    \\
D2     &    $\ \ 9.5 - 10.0$    &        \ \ 9.71           &  0.83     & \ \ 957    \\
D3     &    $   10.0 - 10.5$    &           10.22           &  0.84     & \ \ 670    \\
D4     &    $   10.5 - 11.0$    &           10.70           &  0.87     & \ \ 329    \\
    \hline
I1     &    $\ge \ \  9.0$      &        \ \ 9.68           &  0.84     &    3218    \\
I2     &    $\ge \ \  9.5$      &           10.03           &  0.84     &    1997    \\
I3     &    $\ge 10.0$          &           10.38           &  0.85     &    1040    \\
I4     &    $\ge 10.5$          &           10.72           &  0.87     & \ \ 370    \\
    \hline
  \end{tabular}
\end{table}


\section{Simulated mock catalogues}
\label{sec:mocks}

We have built two sets of 40 mock VVDS surveys,
starting from
semi-analytic galaxy catalogues obtained by
applying the prescriptions of \citet{delucia_blaizot2007} to the
dark-matter halo merging trees extracted from
the Millennium simulation \citep{springel2005}. The Millennium run contains
$N=2160^3$ particles of mass $8.6\times 10^8~h^{-1}~M_\odot$ in a cubic box of
size $500h^{-1}~Mpc$. The simulation was built with a $\Lambda$CDM
cosmological model with $\Omega_m=0.25$, $\Omega_\Lambda=0.75$, $\sigma_8=0.9$
and $H_0=73km~s^{-1}~Mpc^{-1}$.
For details on the semi-analytic model, we refer to
\citet{delucia_blaizot2007} and reference therein. Note that this model uses
the \citet{bc2003} population synthesis model and a \citet{chabrier2003}
Initial Mass Function (IMF) to assign luminosities to model galaxies. Our
mass estimates are based on the use of the same population synthesis model
and IMF \citep{pozzetti2007}.

The two sets of 40 $1\times1$ deg$^2$ light cones were generated with the code
MoMaF \citep{blaizot2005} for 40 independent lines of sight.
The first set is complete in redshift up to $z\sim1.7$ and contains all
galaxies irrespective to their apparent magnitude or stellar mass, 
down to the simulation limit (that corresponds roughly to a
stellar mass of $10^8~M_\odot$).
We refer to these as the {\it full} mocks. They will be used to quantify the
stellar mass incompleteness and its effect on clustering measurements.
Nevertheless, these mocks can not be used to built {\it observed} VVDS-like
mock catalogues because do not include galaxies above $z\sim1.7$.
On the other hand, the second set is complete in 
apparent magnitude up to $I_{AB}=24$
and was generated independently to the first one.
We refer to these mock catalogues as the {\it I24} mocks and will use
them to quantify statistical errors on clustering measurement, using the
full spectroscopic VVDS-02h observing strategy.

The Millennium simulation is particularly appropriate as a test-bench
for the VVDS data, as it is able to reproduce a number of basic properties
of the data, in particular, the number counts in various bands.
Additionally, 
when the VVDS selection function is applied, the average redshift
distribution of the mock samples is in very good agreement with the
VVDS. 

\begin{figure}
  \includegraphics[width=9cm]{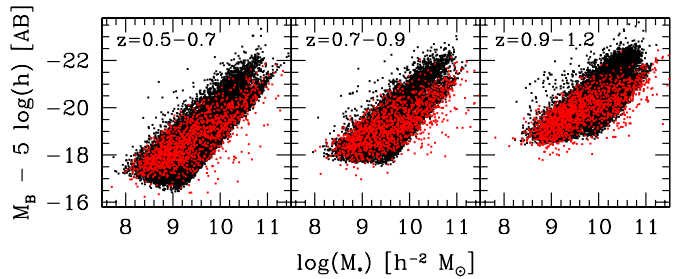}
  \caption{Comparison of the stellar mass vs. rest-frame B band relation in 3
    redshift ranges between mock catalogues (black points) and VVDS data (red points).}
      \label{fig:mass_luminosity}
\end{figure}

Most importantly for the analysis of this paper, the
mass-luminosity relation of the simulated samples matches 
fairly well that of the data.
This is shown in Fig.~\ref{fig:mass_luminosity} in three redshift bins
within the studied range. The general agreement between the envelopes of
the observed and simulated relationships is quite encouraging, given 
the complexity and approximated nature of the semi-analytic models.
There are clearly some differences, which will have to be taken into
account when using the simulation to interpret some of the effects
that are present in the data. The most evident one is an apparent excess of
massive and luminous objects in the mock catalogues, in particular in
the most distant redshift bin. 
This will not affect our
conclusions when using the simulations to understand the level of
incompleteness in mass due to the flux limit of the survey. This test
depends on the objects with high mass-to-light ratio, near the lower
luminosity limit of the data. From the three panels, we can at least say
that the scatter in the relations for the data and the mocks at these
luminosities is very similar. We have good reasons to hope, therefore, that
the results of our tests based on the mocks will provide realistic
indications on the completeness of the data.
A more detailed comparison between properties of mock catalogues and VVDS
data will be addressed in future papers.

\section{Estimating the real-space correlation function}
\label{sec:xi_technic}

\subsection{Methodology}

The two-point correlation function is
the simplest estimator used to quantify galaxy clustering, being
related to the second moment of the galaxy distribution, i.e. its
variance.  In practice, it describes the excess probability to
observe a pair of galaxies at given separation, with respect to that
of a random distribution \citep{peebles1980}. Here we shall use the
redshift-space correlation function \xirppi\ . In this case, galaxy
separations are split into the tangential and radial components,
$r_p$ and $\pi$, as account for redshift-space
distortions \citep{dp83}.

The real-space correlation function $\xi_R(r)$ can be recovered by
projecting \xirppi along the line-of-sight, as
\begin{equation}
  w_p(r_p) \equiv 2 \int_0^\infty \xi(r_p,\pi) dy = 2 \int_0^\infty
  \xi_R\left[(r_p^2+y^2)^{1/2}\right] dy
  \label{wpdef}
\end{equation}

For a power-law correlation function, $\xi_R(r)=(r/r_0)^{-\gamma}$,
this integral can be solved analytically and fitted to the observed
\wprp, and find the best-fitting values of the correlation length
\ro\ and slope \gam\ (e.g. \citet{dp83}).  In computing \wprp,
the upper integration limit has, in practice, to be chosen finite, to
include all the real signal without adding extra noise (which is
dominant above a certain $\pi$).  In our previous works, we have
performed extensive experiments 
\citep{pollo2005}, and found that, for the same data, an integration
limit of $\pi_{max}=20$~h$^{-1}$Mpc provides the best compromise in
terms of noise and systematic bias of the result.  Using 30 or
40~h$^{-1}$Mpc does not change the recovered \wprp\ appreciably,
while increasing the scatter.
Finally, fitting \wprp\ for separations $<10$ $h^{-1}$ Mpc using the
procedure described in detail in \citet{pollo2005}, that takes into account
the covariance matrix of the data and in particular the fact that the bins
are not fully independent, provides a best-fitting value of \ro\ and
\gam\ for each mass-selected sub-sample.

To measure in practice \xirppi\ from each data sample, we use
the standard estimator of \citet{lansal1993}:
\begin{equation}
  \xi(r_p,\pi) = \frac{N_R(N_R-1)}{N_G(N_G-1)} \frac{GG(r_p,\pi)}{RR(r_p,\pi)}
  - \frac{N_R-1}{N_G} \frac{GR(r_p,\pi)}{RR(r_p,\pi)} + 1
  \label{lseq}
\end{equation}
where $N_G$ is the mean galaxy density (or, equivalently, the total
number of objects) in the survey; $N_R$ is the mean density of a
catalogue of random points distributed within the same survey volume and with
the same selection function as the data;
$GG(r_p,\pi)$ is the number of independent galaxy-galaxy pairs with
separation between $r_p$ and $r_p+dr_p$ perpendicular to the
line-of-sight and between $\pi$ and $\pi+d\pi$ along the line of
sight; $RR(r_p,\pi)$ is the number of independent random-random pairs
within the same interval of separations and $GR(r_p,\pi)$ represents
the number of galaxy-random cross pairs.

\subsection{Correction of observational biases}

To properly estimate the correlation function from the VVDS data, we need
to correct for the different sampling rate, which on average is 22.8\%
but varies with the position on the sky due to the VIMOS foot-print
and the superposition of multiple passes \citep{lefevre2005_vvds}. To
this end, in \citet{pollo2005} we developed and tested a 
correction scheme based on computing a specific local weight
around each galaxy with measured redshift. The weight is computed utilizing
the angular information of the ``missed'' galaxies contained in the survey
parent photometric catalogue, which should have the same redshift distribution
as the spectroscopic sample.
In our specific case, in which we are considering the clustering of
galaxies within a given mass range, the parent photometric catalogue
should contain only galaxies in the same range. This information is a
priori not available, as to compute it one needs to know galaxy
distances.  As a remedy to this, we used the photometric redshifts
that have been measured to very good accuracy
($\sigma_{\Delta z/(1+z)}=0.029$ for our redshift range), over the whole
VVDS-02h Deep field \citep{ilbert2006}, thanks to the combination of CFHT-LS
and VVDS photometry. This uncertainty on the galaxy distance corresponds
to an error on the computed stellar mass of $\sim$0.16dex
\citep{pozzetti2007}.
The chosen redshift range
is large enough to assume that edge effects due to photometric
redshift errors are marginal. We stress that here we are not using
the photometric redshifts directly into the correlation estimate,
but only in the correction weight, to assign the galaxy to the
appropriate (broad) redshift slice and mass interval. We used this
information to build, for each mass-selected spectroscopic data set,
a corresponding mass-selected parent photometric catalogue
containing galaxies in the range $0.5\le zphot\le 1.2$. This method
allowed us to simplify the weight expression originally used for the
overall flux-limited sample \citep[equation 12,][]{pollo2005}, such
that the the $i$-th galaxy will have a weight

\begin{equation}
  w_i=\sqrt{\frac{n_p(i)(n_p(i)-1)}{n_s(i)(n_s(i)-1)}}\,\,\,\,   .
  \label{eq:newweight}
\end{equation}

In this expression, $n_p$ is the total number of galaxies in the
parent catalogue inside a circle of fixed radius [chosen to be
40$^{\prime\prime}$ on the basis of mock experiments \citep{pollo2005}],
centred on the $i$-th galaxy, while $n_s$ is the number of
spectroscopically-observed galaxies inside the same area.
The idea behind this weighting scheme is that a pair of galaxies
will contribute to the correlation function with a total weight
$w_iw_j$, proportional to the ratio of the expected-to-measured
numbers of pairs within the chosen radius.

Since the sub-samples analyzed in this work are essentially
volume-limited (above the mass completeness limit discussed in the
next section), we do not need to apply any other weighting scheme as
usually done for purely flux-limited surveys [as e.g.~in \citet{li2006}].

\subsection{Systematic and statistical errors on correlation estimates}

To test for biases in our correction scheme and to quantify 
statistical errors on the estimates of \wprp, we used the 40
{\it I24} mocks catalogues described in Sect.~\ref{sec:mocks}. From the full
$1\times 1~deg^2$ light-cones, we created 40 fully realistic VVDS-02h
surveys, by applying the entire selection function and observing strategy of
the real data as described in \citet{pollo2005}. For each of the 40 cones, 
therefore, we had the possibility to measure both the
``true {\it $I_{AB}\le 24$}'' and ``observed'' values of \wprp\ and the
corresponding power-law-fit parameters to the spatial correlation
function. Comparing the mean and {\it rms} of the difference in the measured 
parameters, we find that (on the average, over the four mass
ranges), the recovered value of \ro\ is 5\% accurate
(the maximum systematic effect found in the case of the I1-like set of
mocks, is 10\%);
similarly, the slope \gam\ is consistent with no
systematic bias, and a scatter of 4\%.
These conclusions are similar to those in \citet{pollo2005}.
Finally, the root-mean-square variation of \wprp\ among the 40 ``observed''
mocks provides us with an estimate of the error bars on our
measurements of \wprp, \ro\ and \gam. Assuming the
simulations are a realistic representation of reality, these error
bars also include cosmic variance from fluctuations
on scales larger than the sample size.

\section{Effects of mass incompleteness}
\label{sec:mass_comp}

The VVDS flux limit in the $I$ band translates at different redshifts
into different lower luminosity limits.  This, in turn, translates
into a broad mass selection cut at each redshift, reflecting the
scatter in the Mass-Luminosity relation. If not treated appropriately,
this introduces a bias against high mass-to-light (ML) ratio galaxies,
i.e. objects that would enter the sample if this were purely mass
selected, but that are excluded simply because they are too faint to
fulfill the apparent magnitude limit.  Furthermore, at a fixed
luminosity in the observed band, it will be the redder objects
(typically more clustered), dominated by low-luminosity stars, to have
the highest ML ratios and thus to be preferentially excluded
(see Fig.~\ref{fig:missed_galaxies}).
If this is not accounted for in
some way, it will inevitably affect the estimated clustering
properties, with respect to a truly complete, mass-selected sample.
It is therefore necessary to understand in detail the effective
completeness level in stellar mass of the samples that we have defined
for our analysis. We have explored this issue using two complementary
and independent approaches based respectively on the data themselves
and on the mock samples.
The latter will also allow us to test directly the effects of the
incompleteness on the measured clustering; the results of this test will be
presented in Sect.~\ref{sec:massincomp_and_wprp}.

\subsection{Observed scatter in the mass-luminosity relation}

\begin{figure}
  \includegraphics[width=9cm]{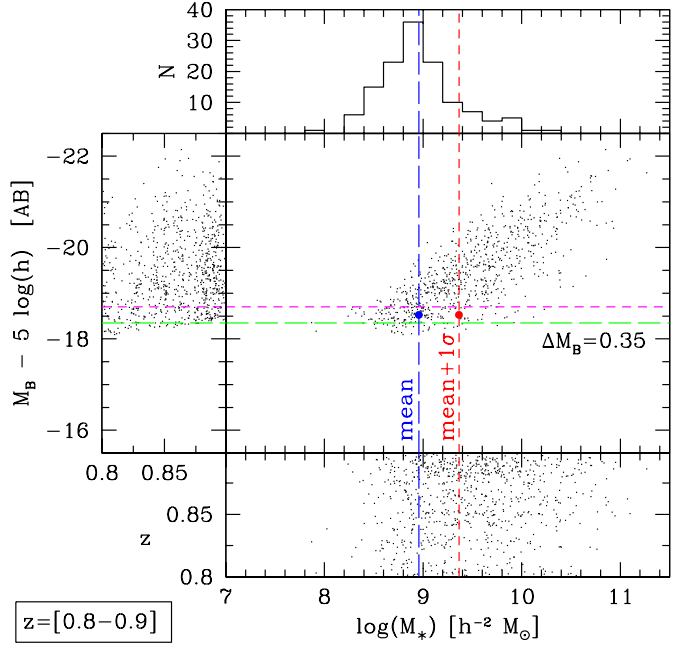}
  \caption{Construction of the completeness limit from VVDS data in
    the narrow redshift range z=[0.8-0.9]. The central panel
    shows the value of B-band luminosity vs. the computed stellar mass
    for all galaxies in that redshift range. The long-dashed
    horizontal line defines the minimum galaxy luminosity above which
    stand 95\% of the galaxies. The top horizontal dashed line defines
    the upper limit in which we compute the mean and standard
    deviation of the stellar mass distribution of galaxies (top panel)
    standing in a luminosity interval of $\Delta M_B=0.35$. The
    right-hand dashed vertical line defines the 84\% completeness
    limit in stellar mass.
  }
  \label{fig:incompleteness_def}
\end{figure}

\begin{figure}
  \includegraphics[width=9cm]{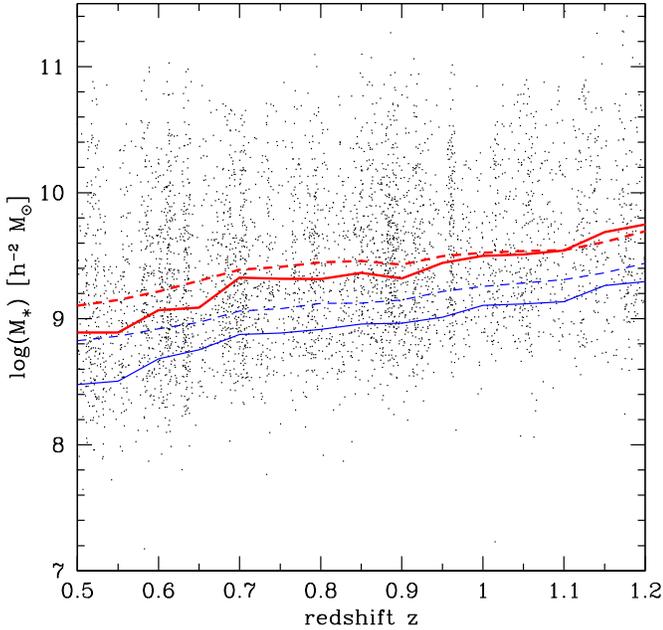}
  \caption{Distribution of stellar masses as a function of redshift,
    in the range [0.5-1.2]. The top (bold) solid line, computed from the
    observed scatter in the luminosity-mass relation, defines the
    limit above which samples are better than 84\% complete. The
    bottom line defines the 50\% completeness limit.
    The two dashed lines represent the same limits but derived from the
    Millennium simulation.
  }
  \label{fig:m_vs_z}
\end{figure}

A first way to understand the incompleteness in mass of our
samples is to measure directly the scatter in the mass-luminosity
relation near the flux limit of the survey and from this estimate
the fraction of missing objects, at the given mass limit.  Since the
flux limit of the survey translates into a different luminosity
limit at different redshifts, we need to do this in narrow redshift
ranges as to track the change of the incompleteness within the broad
redshift range of the sample (and also take into account any
potential evolution in the Mass-Luminosity relation).
The method is illustrated by Fig.~\ref{fig:incompleteness_def} where we plot
the value of B-band luminosity vs. the computed stellar mass for all
galaxies in the redshift range z=[0.8-0.9].
The ``completeness'' limit, defined as the value of stellar mass
above which virtually all masses, given the observed distribution,
are included in the sample is estimated as follows:
1) given a redshift bin ($\Delta z=0.1$), we compute the minimum galaxy
luminosity detected at that distance, defined as the value of absolute
magnitude above which we have 95\% of the galaxies, $M_{min}$; this is
given by the long-dashed horizontal line of Fig.~\ref{fig:incompleteness_def}.
2) We then consider a luminosity interval $[M_{min} ; M_{min}-0.35]$, chosen
as the minimum thickness to provide us with statistics sufficient to define
the properties -- mean and scatter -- of the (logarithmic) mass distribution,
as indicated by the short-dashed horizontal line in the same figure.
We shall define for a specific redshift the completeness limit to be:
\begin{equation}
Y_c(z)=\left<Y(z)\right>+\sigma_Y(z)\,\,\,    ,
\end{equation}
where $Y(z) = \log M(z)$. Note that the completeness limit defines, as a
function of $z$, the value of mass above which, given the selection properties
of the survey, a sample will be better than 84\% complete in mass
(corresponding to one-sided $1\sigma$ limit). 
Note that here we do not have enough data to perform a refined Gaussian
reconstruction as done by \citet{li2006} for the SDSS. The completeness locus
as a function of redshift is plotted as the bold (red) solid line in
Fig.~\ref{fig:m_vs_z} in the plane redshift-mass. Figure~\ref{fig:m_vs_z}
indicates that any sample limited to masses larger than $10^{9.5} M_\odot$
will be fairly complete in stellar mass over the whole redshift range we plan
to explore.  This completeness limit is consistent with the one derived
by \citet{pozzetti2007} in the redshift range [0.9-1.2] in the
determination of the galaxy stellar mass function from the VVDS data (see
their Fig.~9).

\subsection{Mass incompleteness from the mock catalogues} 

Encouraged by the realistic appearance of the Mass-Luminosity
relation for our mock samples (Fig.~\ref{fig:mass_luminosity}), we
can try and use them to independently test the consistency of the
incompleteness level estimated in the previous section.

As a first test, we can apply to the mock samples the direct technique
used on the data in the previous section. The results of this, averaged over
the 40 mock samples, are shown in terms of a completeness curve in the plane
$z-M$ in Fig.~\ref{fig:m_vs_z} (dashed lines). Interestingly, we see that 
the 84\% completeness level obtained from the simulations is very
close to that from the data (upper curve), although the mean values
of the two mass-luminosity relationships are slightly different.

Then, let us exploit the {\it full} mock catalogues defined in
Sect.~\ref{sec:mocks}, which are  
deeper than the magnitude limit of the VVDS and include galaxies with
stellar mass down to $10^9~M_\odot$. It is therefore
possible to quantify independently the fraction of galaxies we miss
when applying the VVDS selection function, provided that
the Mass-Luminosity relation and its scatter are comparable in
the data and in the simulations. We have seen from
Sect.~\ref{sec:mocks} that this is quite a reasonable assumption.

We define, for a given mass selection, the completeness $C$ as the
fraction of objects recovered when the apparent magnitude selection
$17.5\le I_{AB}\le 24$ is applied, with respect to the pure mass-selected
sample. The average values of $C$ among the 40 {\it full} mocks catalogues
are reported in Tab.~\ref{tab:completeness2} for z=[0.5-1.2].
\begin{table}
  \caption{Mean fraction of galaxies in the redshift range z=[0.5-1.2] for
    which $17.5\le I_{AB}\le 24$ among 40 mocks catalogues (serie S1) which
    are complete in stellar mass down to $10^9~M_\odot$, built from the
    Millennium simulation with semi-analytical model of
    \citet{delucia_blaizot2007}}
  \label{tab:completeness2}
  \centering
\begin{tabular}{c c}
\hline
\hline
  $\log(M/M_\odot)$    &   C   \\
\hline
 $\ \ 9.0 - \ \ 9.5$  & 0.37  \\ 
 $\ \ 9.5 -    10.0$  & 0.70  \\
 $   10.0 -    10.5$  & 0.93  \\
 $   10.5 -    11.0$  & 0.99  \\
\hline
     $\ge \ \ 9.0$    & 0.56 \\
     $\ge \ \ 9.5$    & 0.79 \\
     $\ge    10.0$    & 0.94 \\
     $\ge    10.5$    & 1.00 \\
\hline
\end{tabular}
\end{table}

\begin{figure}
  \includegraphics[width=9cm]{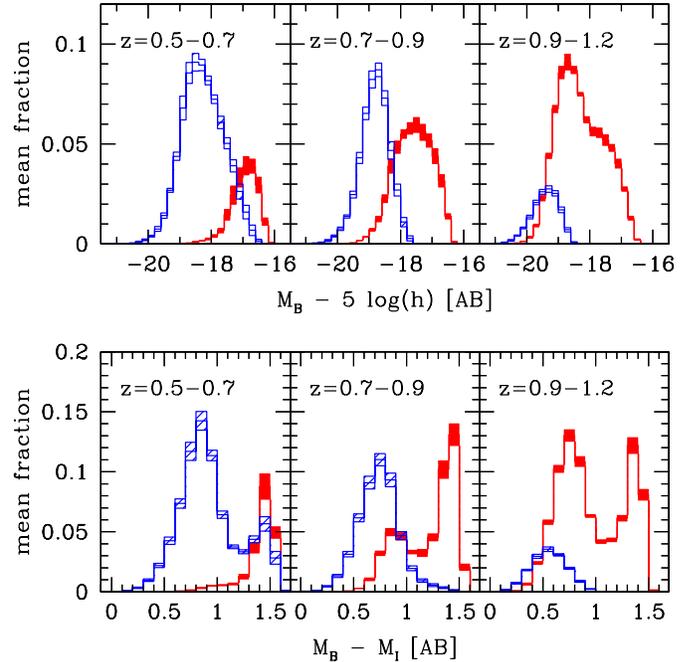}
  \caption{
    Mean luminosity and color distributions, with their
    $1\sigma$ variance, for galaxies with stellar masses in the
    range $\log(M/M_\odot)=[9-9.5]$ in the {\it full} millennium mock
    catalogues, for 3 consecutive redshift ranges between $z=0.5$ and
    $z=1.2$. Solid red distributions correspond to galaxies fainter
    than $I_{AB}=24$, while hatched blue distributions correspond to
    galaxies brighter than this threshold. The distribution have
    been normalized to the total number of galaxies in each redshift
    range. The galaxies missed due to the $I_{AB}=24$ flux limit
    function are clearly intrinsically fainter and relatively
    redder.
  } 
  \label{fig:missed_galaxies}
\end{figure}

These values indicate that for $9<\log(M/M_\odot)<9.5$, our sample
is strongly incomplete, with an expected value of lost galaxies
larger than 60\%. This represents an average value over the
z=[0.5-1.2] redshift range (while we have seen from the previous
analysis that the incompleteness is actually a function of
redshift). More encouragingly, the tests also show that an integral
sample with $\log(M/M_\odot)>9.5$ is already $\sim80\%$ complete and
that samples above $\log(M/M_\odot)=10$ are better than 93\% complete
both in the differential and integral cases.
It is interesting to use the simulated sample to explore which are the
properties of the ``missed'' galaxies. Figure~\ref{fig:missed_galaxies}
shows the luminosity and rest-frame (B-I) color distributions of galaxies
fainter than $I_{AB}=24$ but with masses in the range
$\log(M/M_\odot)=[9.0-9.5]$, compared to those brighter than this limit (and
thus retained within the sample).
Interestingly, the population missed by the VVDS selection function
-- at least for the mock samples
-- is fairly red galaxies, as one would expect for
higher-than-average mass-luminosity ratio objects dominated by old
stars.

Finally, considering the {\it differential} 84\% completeness limit in
Fig.~\ref{fig:m_vs_z}, we note that -- once averaged over
the redshift range -- it roughly corresponds to a mass limit
$\log(M/M_\odot)>[9.5-10]$.  This is consistent with the integral
completeness fraction $C$ derived in this section using the simulations,
for the same range of masses (Table \ref{tab:completeness2}).


\subsection{Effects on the measured clustering}
\label{sec:massincomp_and_wprp}

We can use the 40 {\it full} mock catalogues to go one step
further and
directly quantify the effects of the mass incompleteness on the clustering
measurements, when compared to purely mass-selected samples.

\begin{figure*}
  \includegraphics[width=\textwidth]{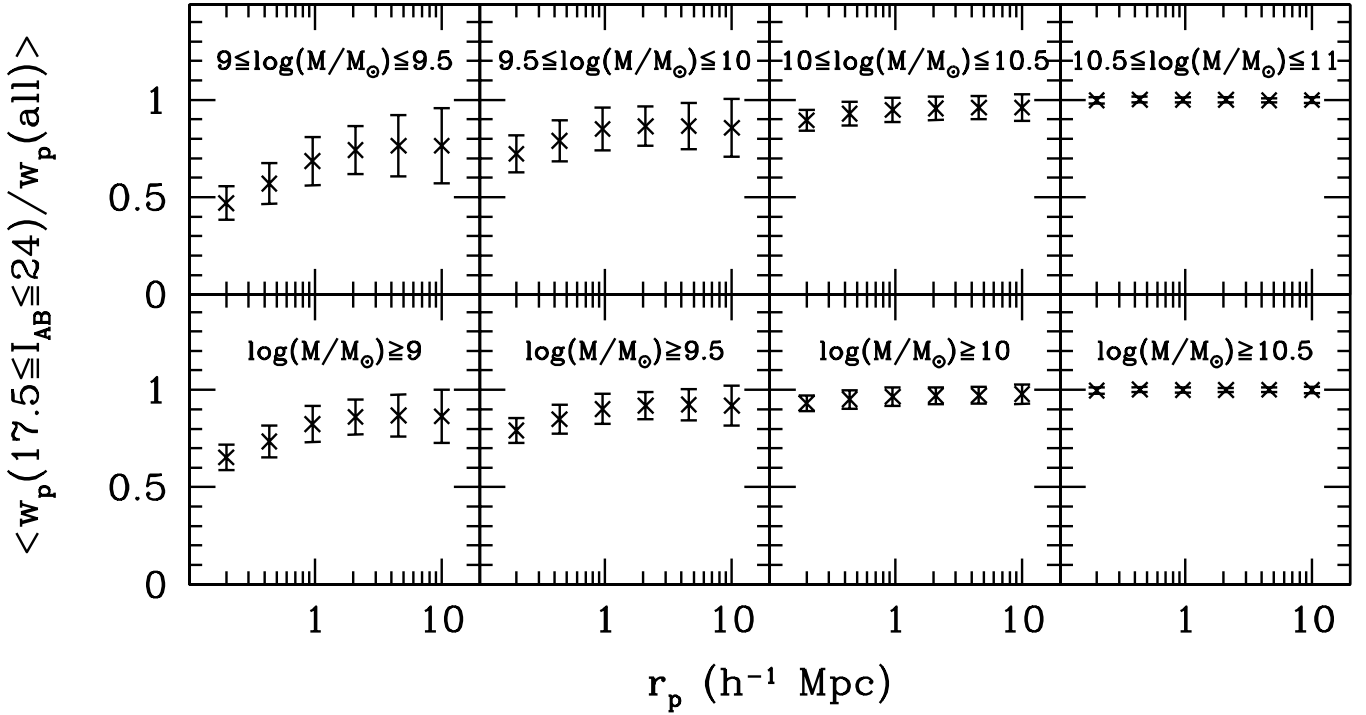}
  \caption{Effect of mass incompleteness on the projected correlation
    function \wprp\ estimated using the Millennium mock catalogues
    for the 8 mass ranges . 
    The points give at each scale $r_p$ the mean ratio of the
    correlation functions \wprp\ measured for a sample complete in
    stellar mass and its corresponding sub-sample selected in apparent
    magnitude $17.5\le I_{AB}\le 24$. Error bars correspond to the
    {\it rms} scatter among the mocks.
  }
  \label{fig:wp-test-mocks}
\end{figure*}

Figure~\ref{fig:wp-test-mocks} shows the average and scatter
over the 40 mocks of the ratio of the projected functions measured
respectively with and without the VVDS flux limit, within the usual
mass ranges.
As expected, the values of \wprp\ are significantly affected by
the mass incompleteness, in particular at very small separations. 
For samples containing masses below $\log(M/M_\odot)\sim10$, not only the
amplitude is underestimated by 15\% or worse, but also the {\it  shape} of
\wprp\ is affected, with the average ratio becoming smaller and smaller
below $1~h^{-1}~Mpc$.
This can be translated in terms of a systematic bias in the
measurements of \ro\ and \gam. The results of averaging the
relative difference $\delta x / x = (x_{true}-x_{obs})/x$ over the
40 mocks, are reported in Tab.~\ref{tab:roga_incompleteness} in
terms of a relative bias $\epsilon(x) = <\delta x / x>_{40}$ on each
parameter. In all cases the mass incompleteness produces a
systematic underestimate of both the amplitude and slope of
$\xi(r)$.

In particular, we have a systematic $\sim25\%$ underestimate of
the correlation length \ro\ in the lowest-mass sample
($\log(M/M_\odot)=[9.0-9.5]$). The slope \gam\ is less affected
(7\%); this is related to the fact that the fit is performed out to
$r_p=10$~h$^{-1}$~Mpc, and thus is not so sensitive to the
small-scale flattening that is evident in Fig.~\ref{fig:wp-test-mocks}. We
also see that the systematic effect is significantly less severe for the
{\it integral} samples (e.g.~only 6\% on \ro\ and 3\% on \gam\ for
$\log(M/M_\odot)\ge9.5$).   

On the basis of these extensive tests, we shall consider as
reliable in the analysis of the VVDS data only results for
samples with a minimum mass limit of $\log(M/M_\odot)=9.5$, with some
caution also in interpreting the results from the sample D2
($9.5<\log(M/M_\odot)<10$), which is still affected by some residual
incompleteness.

In the next section, we shall present the measured values in
general {\it uncorrected} for these effects, but will discuss and
show how they would change assuming a correction for the
incompleteness as derived in Tab.~\ref{tab:roga_incompleteness}. All plotted
error bars on the measured points will correspond to the statistical
uncertainty only, as estimated from the variance in the 40 {\it I24} mock
catalogues.

\begin{table}
  \caption{Average systematic underestimate of
    the clustering length \ro\ and the slope \gam\ of the
    correlation function introduced by the mass incompleteness. The
    values are computed for each parameter as
    $\epsilon(x)~=~<\delta x~/~x>_{40}$, where
    $\delta(x)~=~x_{true}-x_{observed}$ and the
    average is performed over 40 mock samples.  
  }
  \label{tab:roga_incompleteness}
  \centering
  \begin{tabular}{ccc}
    \hline
    \hline
  $\log(M/M_\odot)$    &   \multicolumn{2}{c}{average underestimate}    \\
                      &   ~~~~~ $\epsilon(r_0)$   & $\epsilon(\gamma)$       \\
    \hline
 $\ \ 9.0 - \ \ 9.5$  &    ~~~~~  25\%            &          7\%             \\ 
 $\ \ 9.5 -    10.0$  &    ~~~~~  11\%            &          3\%             \\
 $   10.0 -    10.5$  &    ~~~~~   3\%            &          1\%             \\
 $   10.5 -    11.0$  &    ~~~~~   0\%            &          0\%             \\
    \hline
     $\ge \ \ 9.0$    &    ~~~~~  13\%            &          5\%             \\
     $\ge \ \ 9.5$    &    ~~~~~   6\%            &          3\%             \\
     $\ge    10.0$    &    ~~~~~   2\%            &          1\%             \\
     $\ge    10.5$    &    ~~~~~   0\%            &          0\%             \\
    \hline
  \end{tabular}
\end{table}


\section{Results}
\label{sec:results}

\subsection{\wprp\ as a function of mass at $z\sim0.85$}

\begin{figure*}
 \includegraphics[width=\textwidth]{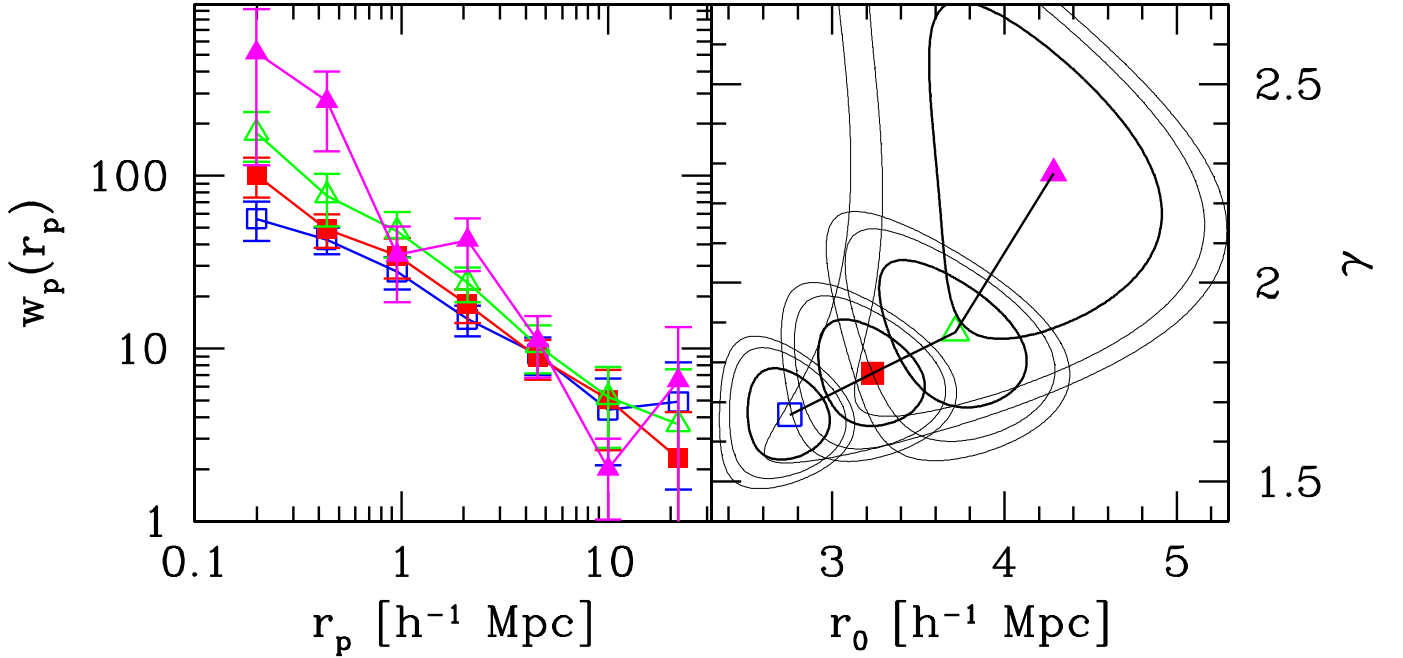}
 \caption{
   {\it (Left)} Measurements of the projected correlation
   function \wprp\ of galaxies with different stellar masses:
   $\log(M/M_\odot)\ge 9.0$ ({\it open blue squares}),
   $\ge 9.5$              ({\it filled red squares}),
   $\ge 10.0$             ({\it open green triangles}) and
   $\ge 10.5$             ({\it filled magenta triangles})
   from VVDS data in the redshift range [0.5-1.2].
   {\it (Right)} The best-fit parameters (\ro\ and \gam) with their
   associated $1-$, $2-$ and $3\sigma$ error contours, derived from the
   variance among 40 mock catalogues.
   In this plot, no correction for the effect of
   incompleteness in mass is applied (see text). Note however that
   the correction would affect both \ro\ and \gam\ and, for
   example, move the open square (the most affected one) in the
   right panel without modifying the observed trend.
 }
  \label{fig:wp_r0_ga}
\end{figure*}

Figure \ref{fig:wp_r0_ga} shows the projected correlation function
\wprp\ computed for the four samples I1 to I4 
(left panel) and its best fit parameters 
together with their $1-$, $2-$ and $3\sigma$ error contours derived
from the 40 {\it I24} observed mock catalogues (right
panel). The measured values of \ro\ and \gam\ for each sample are
reported for convenience also in Tab.~\ref{tab:measurement}, including
those for the sample D1 for which we know that the incompleteness is
severe.
No correction for incompleteness is applied to the
plotted values

\begin{figure}
 \includegraphics[width=9cm]{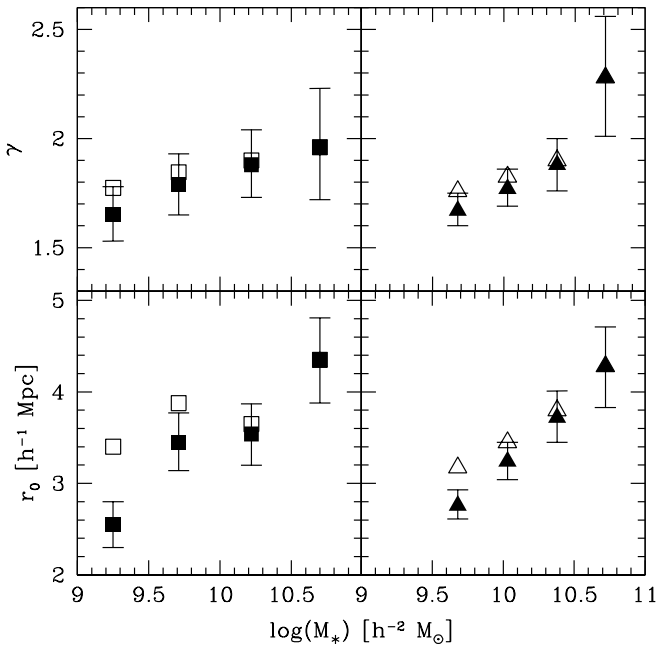}
  \caption{Correlation length and slope measured from VVDS data at
    $z\sim0.85$ as a function of the median stellar mass of each
    samples.
    Left and right panels refer respectively to the
    {\it differential} and {\it integrated} samples. Filed symbols
    indicate measurement non-corrected by stellar mass
    incompleteness, while open symbols are.
  }
  \label{fig:roga_vs_mass}
\end{figure}

Even accounting for the effect of incompleteness and the size of
the error contours, we observe a clear increase of the amplitude and
the slope of the correlation function for increasing median stellar
mass, even if error contours are quite large in the highest mass
range where the statistics is poor (Tab.~\ref{tab:prop}).
This trend is also quite clear in Fig.~\ref{fig:roga_vs_mass}
where instead we have used the simulation results to correct (open
symbols) the observed values (filled symbols), dividing them by
  $(1-\epsilon(x))$ (Tab.~\ref{tab:roga_incompleteness}).  
Note how especially for the integral samples (right panels), the trend with
increasing mass is quite robust and significant. 
We thus conclude that already at $z\sim0.9$ higher mass galaxies 
are significantly more clustered than lower mass objects, with a
systematic trend both in the amplitude and slope of the spatial
correlation function.  
This dependence on stellar mass is coherent in general terms with the
trend found as a function of luminosity \citep{pollo2006}.

\begin{table}
  \caption{Clustering length and slope of the correlation function \wprp\ for
    the different samples in the redshift range z=[0.5-1.2]. Values are
    derived for $0.2\le r_p\le 21~h^{-1}.Mpc$. Associated errors
    have been extrapolated from the variance measured among 40 mocks
    catalogues for each stellar mass range. Values have not been corrected by
    underestimate due to stellar mass incompleteness.
  } 
  \label{tab:measurement}
  \centering
  \begin{tabular}{c c c c}
    \hline\hline
    Sample &    $\log(M/M_\odot)$     & \ro\ ($h^{-1}.Mpc$)     & \gam \\
    \hline
    D1     &    $\ \ 9.0 - \ \ 9.5$  &  $2.55_{-0.25}^{+0.25}$   &   $1.65_{-0.12}^{+0.13}$ \\
    D2     &    $\ \ 9.5 -    10.0$  &  $3.45_{-0.31}^{+0.32}$   &   $1.79_{-0.14}^{+0.14}$ \\
    D3     &       $10.0 -    10.5$  &  $3.54_{-0.34}^{+0.33}$   &   $1.88_{-0.15}^{+0.16}$ \\
    D4     &       $10.5 -    11.0$  &  $4.35_{-0.47}^{+0.46}$   &   $1.96_{-0.24}^{+0.27}$ \\
    \hline
    I1     &    $\ge~~9.0$           &  $2.76_{-0.15}^{+0.17}$   &   $1.67_{-0.07}^{+0.08}$ \\
    I2     &    $\ge~~9.5$           &  $3.24_{-0.20}^{+0.21}$   &   $1.77_{-0.08}^{+0.09}$ \\
    I3     &    $\ge10.0$            &  $3.72_{-0.27}^{+0.29}$   &   $1.88_{-0.12}^{+0.12}$ \\
    I4     &    $\ge10.5$            &  $4.28_{-0.45}^{+0.43}$   &   $2.28_{-0.27}^{+0.28}$ \\
    \hline
  \end{tabular}
\end{table}

\subsection{Comparison to SDSS}

\begin{figure*}
 \includegraphics[width=\textwidth]{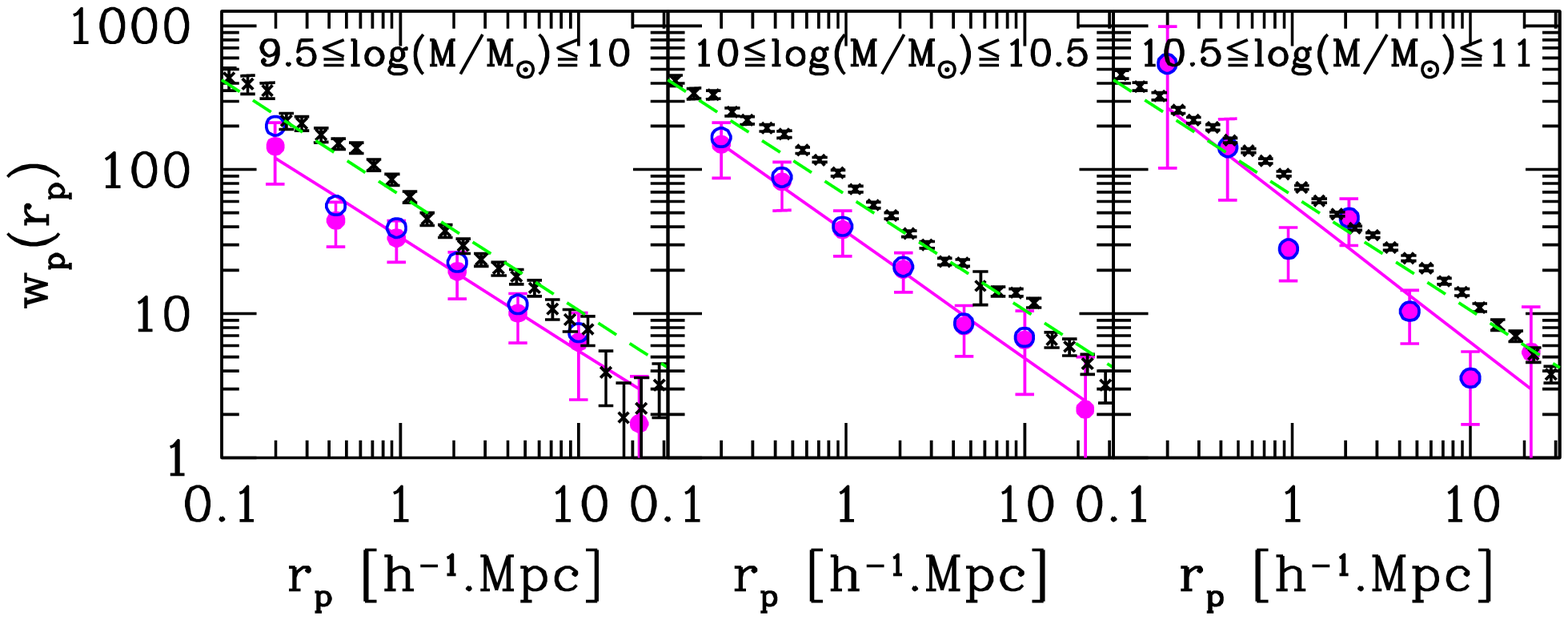}
  \caption{Comparison of the projected correlation function \wprp\
    from SDSS data at z$\sim$0.15 \citep[{\it black cross;}][]{li2006}
    to VVDS measurement at z$\sim$0.85 ({\it magenta filled dots and
      solid lines}) in 3 differential ranges of stellar masses. The
    green dashed line is a power-law reference line drawn with
    $r_0=5$~h$^{-1}$.Mpc and $\gamma=1.8$. Error bars on VVDS
    measurement have been estimated from the variance among 40 mock
    catalogues.
    Blue open circles indicate VVDS measurements corrected
    for stellar mass incompleteness.
  }
  \label{fig:wp_vvds_sdss}
\end{figure*}

This segregation in term of stellar mass at redshift $z\sim0.85$ is
also observed at low redshift in the SDSS data \citep{li2006}. These
authors present their result directly in terms of the measured
amplitude of \wprp\ on different physical scales because of the
departure of the correlation function from a pure power law observed
by numerous authors \citep[e.g.~][]{guzzo1991, zehavi2004}. In
Fig.~\ref{fig:wp_vvds_sdss} we plot our measurements of \wprp\ at
$z\sim0.85$ together with those by \citet{li2006} at $z\sim0.15$
within the three mass ranges in which we have established that the
incompleteness does not strongly affect the measurements.  A reference
power-law \wprp, corresponding to $r_0=5$ h$^{-1}$ Mpc and
$\gamma=1.8$ is over-plotted for reference. 
Blue open circles correspond to correcting the observed values of \wprp\
according to the results of Fig.~\ref{fig:wp-test-mocks}.

Figure~\ref{fig:wp_vvds_sdss} not only shows the dependence of galaxy
clustering on stellar mass at a given redshift but also, at the same
time, the evolution of the amplitude and shape of the correlation
function \wprp\ with redshift for a given stellar mass range. The
observed evolution is apparently faster for low-mass objects
than for massive ones:  
the amplitude of \wprp\ increases by a factor $\sim$2-3.
Conversely, in the high-mass range,
$\log(M/M_\odot)=[10.5-11.0]$, the amplitudes at $z\sim0.85$ and
$z\sim0.15$ are very similar, within the error bars.
These results hold even when considering corrections to the projected
function accounting for the stellar mass incompleteness
(blue empty circles, Fig.~\ref{fig:wp_vvds_sdss}).

\subsection{Comparison to model predictions}
\label{sec:vvdsmillennium}

\begin{figure*}
 \includegraphics[width=\textwidth]{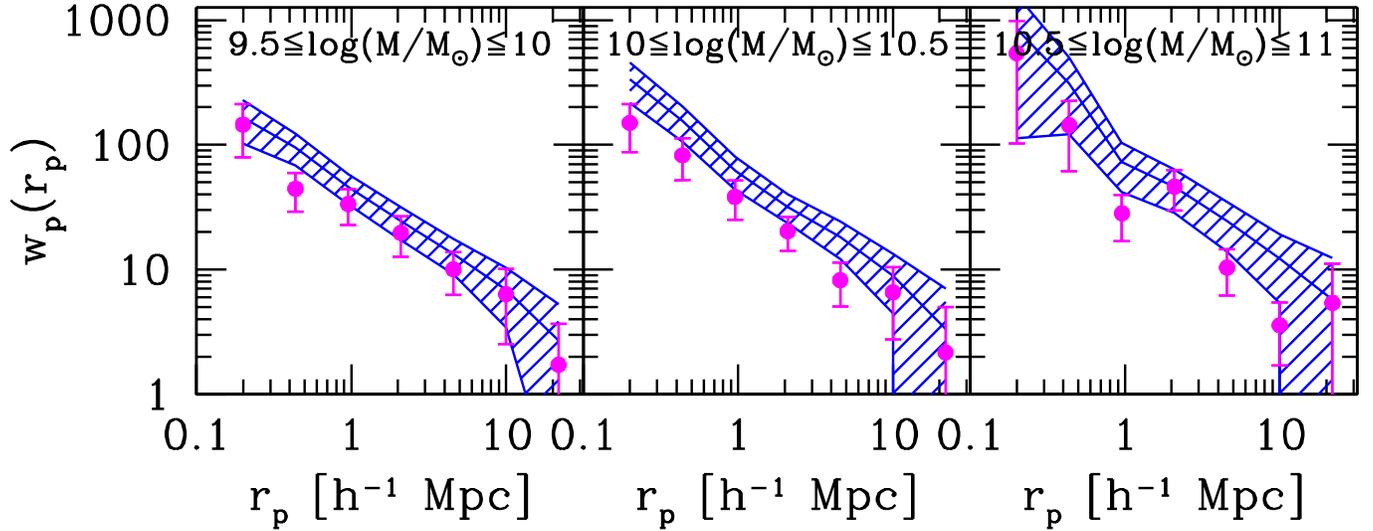}
  \caption{Comparison of the VVDS measurements of the projected
    correlation function \wprp\ as a function of stellar mass
    (filled circles), with that obtained from 40 mock VVDS surveys
    constructed from the Millennium Simulation.  The dashed area
    describes the {\it rms} among the 40 simulated catalogues.  }
  \label{fig:wp_vvds_millennium}
\end{figure*}

So far, we have used the mock VVDS surveys built from the Millennium
Simulation to assess where our results are reliable against the survey
selection effects and established the range of stellar masses where
this is plausibly the case.  In this section we would like to exploit
more fully their intrinsic scientific content.  As discussed
previously, these mock samples were constructed by populating dark
matter halos from the Millennium Run \citep{springel2005}, using the
latest version of the Munich semi-analytic model \citep{delucia_blaizot2007}.
We have shown this model is in quite agreement with a number of observational
results. 
As we already discussed, the ``ideal'' light cones were ``observed'' to
produce realistic replicas of the VVDS-Deep survey, with the same sampling,
field pattern and incompleteness.  They are therefore ideal for providing a
direct comparison, under the same conditions, to the \wprp
measured from the VVDS data.  This is shown in
Fig.~\ref{fig:wp_vvds_millennium}, where the data points are compared
to the average of the measurements from the 40 observed mock
samples (solid line), including a $\pm1 \sigma$ error corridor.  The
agreement of the data and the model predictions for the redshift range
considered, $0.5<z<1.2$, is rather good for all three mass ranges
explored. The VVDS points are consistent with the predictions of the
semi-analytic model within $1 \sigma$ or better. A more comprehensive
comparison of these models with the clustering of VVDS galaxies as a
function of different properties (luminosity, color, spectral type),
will be presented in a separate paper.

Finally, we also note that a similar dependence on stellar mass is
also predicted in the hydrodynamical simulations of
\citet{weinberg2004}, in which the steepening of the correlation
function is particularly marked on scales below $0.5 h^{-1} Mpc$.

\subsection{Evolution on galaxy bias}
\label{sec:bias}

Our results suggest that the apparent evolution of clustering from
$z\sim1$ to the current epoch has been stronger for the less massive
galaxies than for the most massive ones.  In other words, the relative
bias of massive to less-massive galaxies was larger at $z\sim1$ than
it is today. This behavior is also fairly well reproduced by
the Millennium mocks.

In the currently standard scenario of galaxy formation, galaxies form and
evolve inside dark matter haloes \citep{whiterees1978}, with the first objects
appearing inside the most massive of these. These are expected to correspond
to the highest, more rare fluctuations of the overall matter density field.
On simple statistical grounds, it can be shown that these peaks will be more
clustered than the general dark matter distribution, i.e. they will be
{\it biased} \citep{kaiser1984}. Thus, from very simple considerations, if a
larger stellar mass corresponds to a larger hosting dark matter halo mass,
then the most massive galaxies at high redshift will correspond to the highest
peaks and will be highly biased. Strong clustering and high bias are observed
both for massive spheroidals at redshifts in the range $z\sim1.5-2.5$,
selected on the basis of their red optical-infrared colors
\citep[e.g.][]{daddi2003}, and for strongly star-forming galaxies at $z\sim3$
selected from their strong UV Lyman break \citep[e.g.][]{steidel2004}.

Another important piece of information is that, when selected on the
basis of a fixed rest-frame luminosity, galaxies show an increase of
their average bias as a function of redshift.  Using the VVDS data, we
have in particular shown that the bias of galaxies with $M_b\le
-20+5\log{\rm h}$ increases by a factor $\sim1.5$ between $z=0$ and
$z\sim1.5$ \citep{marinoni2005}. However, we also know that the typical
luminosity of the general population of galaxies increases by more than a
magnitude over the same redshift range, which implies that at a fixed absolute
luminosity we are in fact looking at different populations of objects
at different redshifts.  Given that there is clear evidence for at most a
factor of 2 increase in the stellar mass between $z=0$ and $z=1$
\citep{arnouts2007}, the analysis presented here
has the virtue of addressing the dependence of clustering with respect
to a more physical and stable property of the galaxy population.
Even by allowing for a factor of 2 increase in the mass (i.e. assuming
that the progenitors of objects with mass $\bar M$ at $z\sim0$ are
galaxies with mass $\bar M /2$, at $z\sim1$), our results
would not change significantly.

It is therefore of interest to interpret in a more quantitative way
our findings, by estimating the evolution of the linear bias for
galaxies with similar mass at different redshifts.  In
\citet{marinoni2005} we showed that the bias is in fact non-linear, at
about the 10\% level.  This has very little effect on the present
discussion, and we can safely assume here a linear bias model for $b_L$,
where the galaxy {\it rms} fluctuations $\sigma_{R,g}$ on a given scale
$R$ can be expressed as
\begin{equation}
\sigma_{R,g}=b_L \sigma_{R,m}\,\,\,\,  .
\label{eq:bias2}
\end{equation}
with $\sigma_{R,m}$ the mass {\it rms} fluctuations on the same scale.
We also make the further assumption that the bias $b_L$ is independent
of scale, which is very reasonable on sufficiently large scales.  
It is customary to express the {\it rms} fluctuation of galaxies and
mass in spheres of radius $R=8$ h$^{-1}$Mpc, with the value of
$\sigma_{8,m}$ adopted as standard expression for the normalization
of the mass power spectrum at the present epoch. Here we adopt the
value $\sigma_{8,m}=0.76$ \citep{spergel2007}.
This value at $z=0$ can be scaled to the epochs of interest here by
considering the growth of fluctuations in our assumed model,
$\sigma_{8,m}(z)=\sigma_{8,m}(z=0)D(z)$, where $D(z)$ is the linear
growth factor of density fluctuations, $D(z)=g(z)/[g(0)(1+z)]$, and
the normalized growth factor $g(z)$ can be approximated as 
\citep{carroll1992,mowhite2002} 
\begin{equation}
g(z)\approx 
{5\over 2}\Omega_{\rm m}\left[\Omega_{\rm m}^{4/7}-\Omega_\Lambda
+(1+\Omega_{\rm m}/2)(1+\Omega_\Lambda/70)\right]^{-1}\,\,\,\, ,
\end{equation}
with $\Omega_{\rm m}$ and $\Omega_\Lambda$ the density parameters in non
relativistic matter and cosmological constant. Note that in this
expression,
\begin{equation}
\Omega_{\rm m}(z)={\omnow (1+z)^3\over E^2(z)}\, ,
~~~~~
\Omega_\Lambda\equiv\Omega_\Lambda (z)= {\ovnow\over E^2(z)}\,.
\end{equation}
with $\omnow$ and $\ovnow$ their present-day values and
\begin{equation}
E^2(z)=\left[\ovnow+(1-\Omega_0)(1+z)^2
+\omnow (1+z)^3\right]\,\,\,\, ,
\end{equation}
the normalized expansion factor.

The corresponding value of the {\it rms} galaxy fluctuations on the
same scale can be estimated from the parameters of the correlation
function under the assumption of a power-law form,
$\xi(r)=(r_0/r)^\gamma$ \citep{peebles1980}, as
\begin{equation}
\sigma_{8,g} = \sqrt{C_\gamma \left(\frac{r_0}{8
      Mpc/h}\right)}\,\,\,\, ,
\label{eq:bias3}
\end{equation}
where
\begin{equation}
C_\gamma = \frac{72}{(3-\gamma)(4-\gamma)(6-\gamma)2^\gamma}\,\,\,\, .
\label{eq:bias4}
\end{equation}

\begin{figure*}
\includegraphics[width=\textwidth]{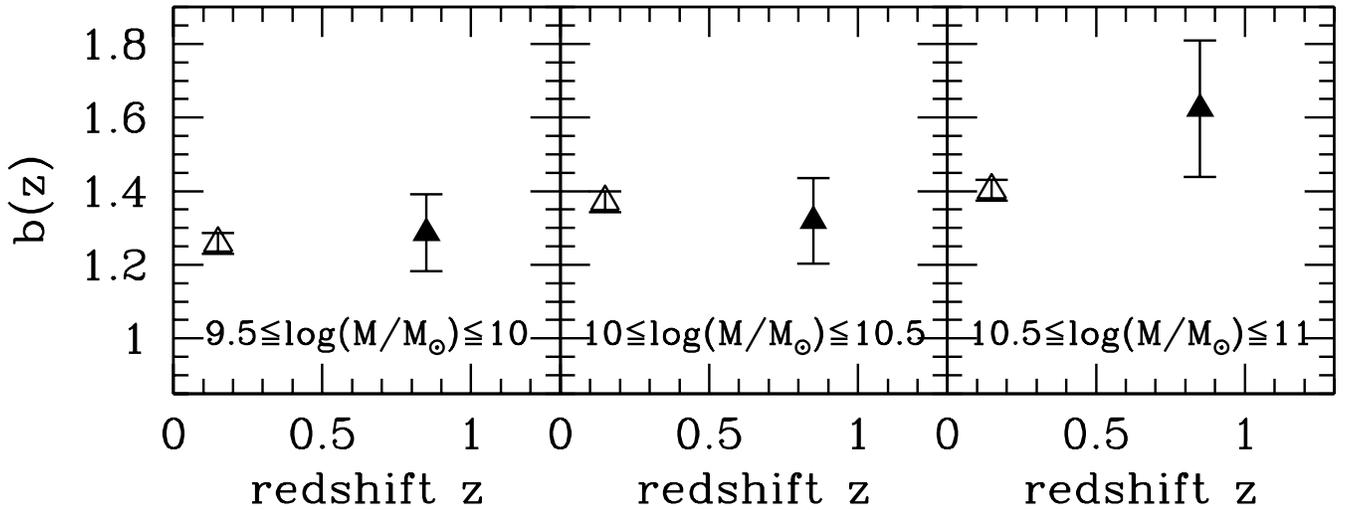}
  \caption{The redshift evolution of the linear bias for galaxies of
    different stellar mass, obtained from the SDSS ($z\sim0.15$, open
    squares, \citep{li2006}), and the VVDS ($z\sim0.85$, filled
    squares, this work). 
  }
  \label{fig:bias}
\end{figure*}

To be consistent with our work at $z=0.85$, we estimated the
parameters \ro\ and \gam\ for the SDSS at $z\sim0.15$ from the published
measurements of \wprp\ by \citet{li2006}, fitting over the same range
used for the VVDS ($0.18\le r_p\le 17.9 h^{-1} Mpc$).
The resulting values of $b_L$ (with $\omnow=0.3$ and $\ovnow=0.7$) over
the three mass ranges and at the two redshifts considered are plotted in
Fig.~\ref{fig:bias}. The values are also reported for reference in
Tab.~\ref{tab:bias}. 

\begin{table}
  \caption{Measured values of the linear bias for the VVDS and SDSS samples,
    within 3 stellar mass ranges (see text).} 
  \label{tab:bias}
  \centering
\begin{tabular}{c c c c}
\hline
\hline
Sample &   $\log(M/M_\odot)$  &          \multicolumn{2}{c}{b(z)}        \\
       &             & SDSS ($z\sim0.15$) & VVDS ($z\sim0.85$)  \\
\hline
  D2   &$\ \ 9.5 -    10.0$ &    $1.26\pm 0.03$  & $1.29\pm 0.10$      \\
  D3   &   $10.0 -    10.5$ &    $1.37\pm 0.03$  & $1.32\pm 0.12$      \\
  D4   &   $10.5 -    11.0$ &    $1.40\pm 0.03$  & $1.62\pm 0.18$      \\
\hline
\end{tabular}
\end{table}

Figure~\ref{fig:bias} shows that for galaxies in the
highest-mass range considered here ($10.5\le \log(M/M_\odot)\le 11$), the
evolution of the linear bias between redshift 0.15 and 0.85 seems
stronger than for the lower mass ranges, where it is substantially
constant over the same redshift range.  For the high mass range, the
increase in the bias going back in time corresponds to a factor
$\sim1.2$ (from $b=1.40\pm 0.03$ to $b=1.62\pm 0.18$)

This kind of behavior is generally expected for the
clustering evolution of dark-matter halos of different mass. 
\citep{mowhite1996,shethmotormen2001,mowhite2002}. More
massive halos form near the peaks of the density field, whose
distribution is strongly biased with respect to the overall mass
distribution and displays a higher clustering amplitude
\citep{kaiser1984}. Our result therefore suggests a direct
relationship (as one would naively expect), between the total mass in
stars accumulated today (or at $z\sim1$) and that of the dark-matter
halo within which the bulk of the baryonic material was originally
assembled.
Clearly, to be still observed at these redshifts, this proportionality needs
to have been preserved also through the merging events experienced by the
galaxy. These produce a merging of the dark-matter halo and a parallel full
coalescence of the baryonic components, with the new baryonic remnant still
sitting in the center of its new, more massive halo. This is not the case when
the galaxy is accreted by a much larger halo (e.g. a cluster), becoming a
satellite whose own halo will be significantly disturbed and modified in mass
by dynamical friction.  This scenario fits well with the results of recent
simulations \citep{conroy2006,wang2006}, that are able to reproduce the
observed clustering properties of galaxies as a function of luminosity at very
different redshifts (up to $z\sim3$) by simply assuming that each dark-matter
sub-halo would generate a galaxy characterized by a B-band luminosity
proportional to its original ``infall'' mass (i.e. mass before being last
accreted into a larger halo).

\section{Summary and conclusions}
\label{sec:conclusion}

We have used the VVDS-Deep data \citep{lefevre2005_vvds} to perform a
first exploration of the dependence of galaxy clustering on
stellar mass at a redshift approaching unity.  This has only now
become possible at these epochs, thanks to the large number of
redshifts and extended wavelength coverage available from VVDS-Deep
that allow us to compute reliable stellar masses.  
We started with a sample of 3218 galaxies 
with $M>10^9~M_\odot$ in the redshift range $z=[0.5-1.2]$. We
thoroughly investigated the completeness limit in mass of the
redshift catalogue, using mock samples built from the
Millennium simulation coupled to semi-analytical recipes.
We have found that there is a significant mass incompleteness induced
by the flux-limit of the survey below $M=10^{10}~M_\odot$, whose
effects on the measured clustering become particularly severe below
$M=10^{9.5}~M_\odot$.  We used the mock surveys to quantify this
systematic effect and to estimate realistic statistical error bars
that possibly include ``cosmic'' variance from fluctuations on scales
larger than the explored volume.  With these limitations in mind, we
have obtained a series of results, that can be summarized as follows:
\begin{itemize}
\item We have measured for the first time a clear stellar mass-dependent
  clustering of galaxies with respect to their stellar mass at a redshift as
  high as $z\sim0.85$, with the more massive objects being more clustered than
  less massive ones.  In particular, the most massive objects show an increase
  in both amplitude and slope of the spatial correlation function.  
\item These clustering properties are very well reproduced (within 1-$\sigma$)
  in the same redshift and mass ranges, by the mock samples built from the
  Millennium run. Together with their realistic redshift distribution,
  this represents a remarkable achievement of the models.
\item Comparison of our measurements to the similar ones obtained
  (with much better statistics) from the SDSS data at $z\sim0.15$
  \citep{li2006} show evidence for a more significant change in the apparent
  clustering amplitude for low-mass galaxies ($\log(M/M_\odot)\le 10$)
  than in the case of the most massive objects ($\log(M/M_\odot)\sim
  11$). The correlation function of the latter, in fact, remains
  roughly constant with time.
\item Assuming a standard cosmology, we computed the expected
  evolution of the root-mean-squared fluctuation in the mass,
  $\sigma_8$, between the two epochs sampled by VVDS and SDSS. Using
  this value, we translated the observed evolution of \wprp\ into
  the corresponding linear bias, showing that the most massive
  galaxies display an evolution of the bias factor, from $b=1.40\pm
  0.03$ at $z\sim0.15$ to $b=1.62\pm 0.18$ at $z\sim0.85$.  This
  finding is not unexpected in a hierarchical scenario in which the
  most massive peaks of the mass density field collapse earlier and
  evolve faster \citep{mowhite1996}.  This interpretation supports a
  scenario in which the stellar mass of a galaxy is essentially proportional
  to the mass of the dark-matter halo in which it was last the central
  object, consistent with recent simulations \citep{conroy2006,wang2006}.

\end{itemize}

The combination of these measurements of clustering as a function of stellar
mass and as a function of luminosity at the same redshift
\citep{pollo2006}, together with the analogous SDSS results at $z\sim
0.15$, can provide very important complementary constraints to models
describing how galaxies are distributed within dark matter halos
\citep{zheng2007}. This can help us to understand the
evolution of galaxies of different mass below $z\sim1$ and in
particular on the role of mergers within this redshift range.


\begin{acknowledgements}
  We thank the anonymous referee for the thorough review of the
  manuscript, which resulted in a significant improvement of the
  paper.
  We thank S. Phleps and D. Wilman for a careful reading of the manuscript. 

  BM thanks the Osservatorio di Brera for the generous hospitality.
  LG thanks S. White and G. Kauffmann for
  discussions and suggestions during the development of this work.
  He also gratefully acknowledges the hospitality of MPE, MPA and
  ESO. BM and LG warmly acknowledge the unique stimulating atmosphere
  of the Aspen Center for Physics, where this work was completed.

  This research program has been developed within the framework of the VVDS
  consortium and has been partially supported by the CNRS-INSU and its
  Programme National de Cosmologie (France), and by Italian Ministry (MIUR)
  grants COFIN2000 (MM02037133) and COFIN2003 (num.2003020150).

  The VLT-VIMOS observations have been carried out on guaranteed time
  (GTO) allocated by the European Southern Observatory (ESO) to the
  VIRMOS consortium, under a contractual agreement between the Centre
  National de la Recherche Scientifique of France, heading a
  consortium of French and Italian institutes, and ESO, to design,
  manufacture and test the VIMOS instrument.

\end{acknowledgements}

\end{document}